\documentclass[journal=jacsat,manuscript=article]{achemso}

\usepackage{chemformula} 
\usepackage[T1]{fontenc} 
\usepackage{graphicx}
\usepackage{dcolumn}
\usepackage{bm}
\usepackage{amsmath}
\usepackage{physics}
\usepackage{multirow}

\let\oldAA\AA
\renewcommand{\AA}{\text{\normalfont\oldAA}}

\usepackage{upgreek}
\usepackage{amsmath}
\usepackage{subfigure}
\usepackage{hyperref}
\usepackage[normalem]{ulem}
\hypersetup{
	colorlinks,
	citecolor=blue,
	filecolor=black,
	linkcolor=blue,
	urlcolor=blue
}
\usepackage{epstopdf}
\usepackage{relsize}
\usepackage[version=3]{mhchem}
\usepackage{balance}
\usepackage{mathptmx}
\usepackage{sectsty}
\usepackage{graphicx} 
\usepackage{lastpage}
\usepackage[format=plain,justification=justified,singlelinecheck=false,font={stretch=1.125,small,sf},labelfont=bf,labelsep=space]{caption}
\usepackage{float}
\usepackage{fancyhdr}
\usepackage{fnpos}
\usepackage[english]{babel}
\addto{\captionsenglish}{%
	
}
\usepackage{array}
\usepackage{droidsans}
\usepackage{charter}
\usepackage[T1]{fontenc}
\usepackage{setspace}
\usepackage[compact]{titlesec}
\usepackage{multirow}

\usepackage{epstopdf}

\definecolor{cream}{RGB}{222,217,201}
\author{Afreen Anamul Haque}
\affiliation{Department of Electronics and Electrical Communication Engineering, Indian Institute of Technology Kharagpur, Kharagpur-721302, India}
\author{Suraj G. Dhongade}
\affiliation{Department of Electronics and Electrical Communication Engineering, Indian Institute of Technology Kharagpur, Kharagpur-721302, India}
\author{Aniket Singha}
\affiliation{Department of Electronics and Electrical Communication Engineering, Indian Institute of Technology Kharagpur, Kharagpur-721302, India}
\email{aniket@ece.iitkgp.ac.in}

\title[An \textsf{achemso} demo]
  {Gas Sensing Properties of Novel  Indium Oxide monolayer: A First-Principles Study}

\abbreviations{IR,NMR,UV}

\begin{document}



\begin{abstract}
  	We present a comprehensive first-principles investigation into the gas sensing capabilities of a novel two-dimensional Indium Oxide (In\textsubscript{2}O\textsubscript{3}) monolayer, using  density functional theory (DFT) calculations. Targeting both resistive-type and work function-based detection mechanisms, we evaluate the monolayer's interactions with  ten hazardous species: NH\textsubscript{3}, NO, NO\textsubscript{2}, SO\textsubscript{2}, CS\textsubscript{2}, H\textsubscript{2}S, HCN, CCl\textsubscript{2}O, CH\textsubscript{2}O, and CO. To assess the deployability of the sensor in ambient environments, we also analyze its interaction with common atmospheric or  background gas molecules, such as, O\textsubscript{2}, CO\textsubscript{2}, and H\textsubscript{2}O. The monolayer exhibits pronounced sensitivity towards NO and H\textsubscript{2}S via substantial conductivity and workfunction modulation, while NH\textsubscript{3} and HCN are readily detected through significant work function shifts only. Biaxial mechanical strain  further proves highly effective in broadening the sensing capability. Tensile strain  adjusts the adsorption energy favourably in most cases and additionaly facilitates  detection of NO\textsubscript{2}  molecule via conductivity modulation. Furthermore, tensile strain  also induces notable work function modulation, enabling detection of additional analytes such as NO\textsubscript{2}, CS\textsubscript{2}, CCl\textsubscript{2}O, and CO   which were undetectable by a work-function based set-up earlier. On the contrary, compressive strain facilitates detection of CH\textsubscript{2}O via workfunction modulation. Although the monolayer remains unresponsive to ambient O\textsubscript{2} and CO\textsubscript{2} molecules,  the moderate adsorption energy of H\textsubscript{2}O may impact selectivity in humid environments. These results establish 2D In\textsubscript{2}O\textsubscript{3} as a highly promising and tunable platform for next-generation miniaturized gas sensors suited for environmental monitoring and safety-critical applications.
\end{abstract}

\tableofcontents
\section{Introduction}
\label{sec:introduction}
\begin{figure}
	\centering
	\includegraphics[width=0.8\textwidth]{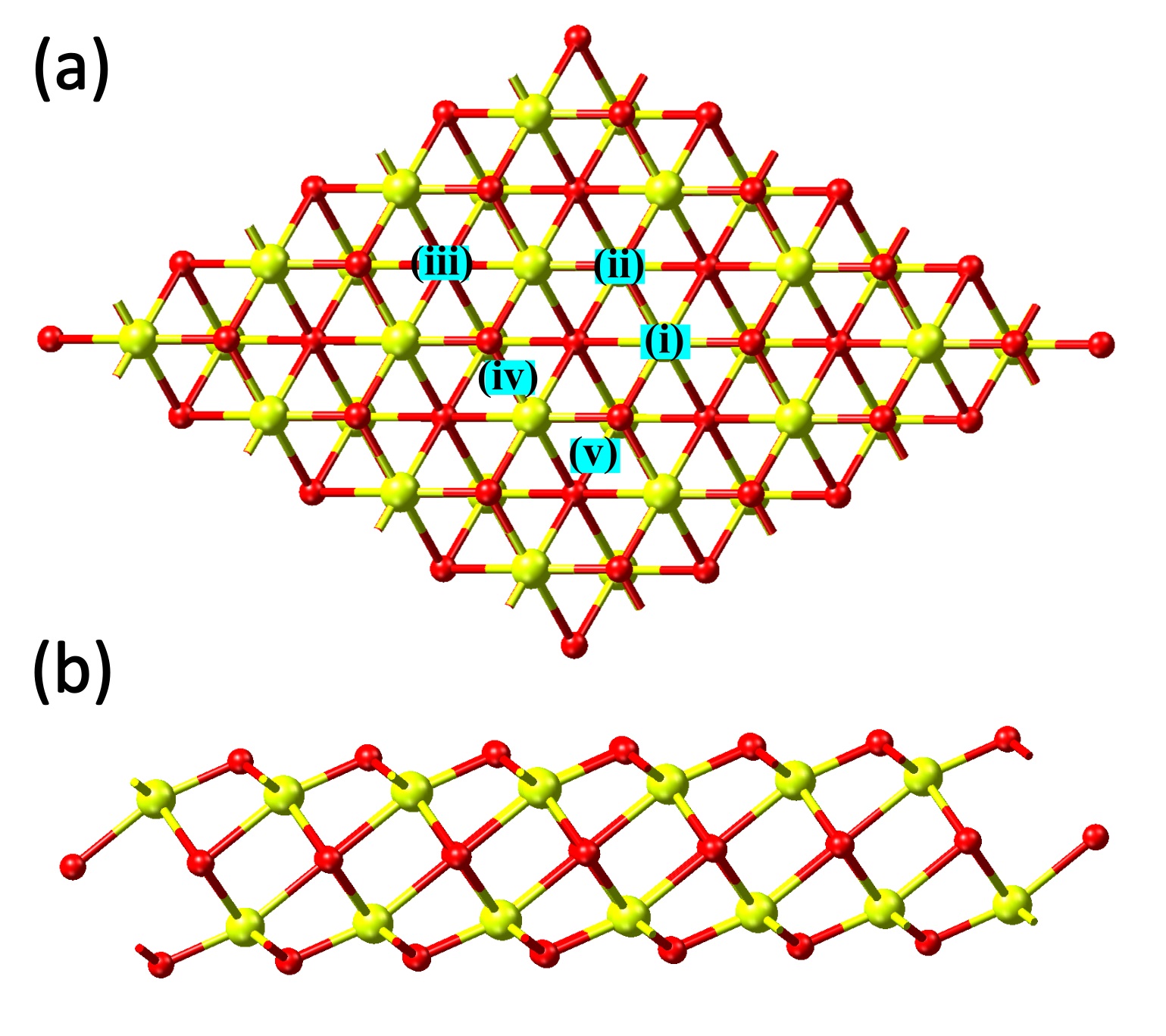}
	\caption{(a) Top and (b) side views of the novel pristine  In\textsubscript{2}O\textsubscript{3} monolayer \cite{in2o3}.The high-symmetry sites considered for possible gas molecule adsorption are labeled with Roman numerals (i) to (v) in panel (a).}
	\label{schematic_monolayer}
\end{figure}

Two-dimensional (2D) materials have emerged as a pivotal platform in contemporary nanotechnology, particularly since the isolation of monolayer graphene. Their growing prominence is attributed not only to their exceptional physical and chemical properties but also to their broad applicability across diverse technological domains. A defining feature of 2D systems is their ultra-high surface-to-volume ratio, which significantly enhances their chemical reactivity and interaction with external species \cite{b1,b2,b3}. Their electronic properties can also be precisely tuned via external stimuli—such as in-plane strain, electric fields, and chemical doping—offering considerable flexibility for device engineering.  An additional advantage of 2D materials is their compositional tunability, which enables phase transitions among insulating, semiconducting, and metallic states through tailored elemental configurations \cite{b4,b5,b6}. This versatility is further augmented by weak interlayer van der Waals forces, facilitating both surface-level and interlayer structural modifications. Such characteristics render 2D materials especially suitable for sensing applications, where attributes like electronic responsiveness, chemical selectivity, and structural adaptability are of paramount importance.\\
\indent Gas sensing, in particular, stands to benefit from the unique confluence of high electrical conductivity, mechanical flexibility, tunable electronic band structures, and functionalization potential inherent to 2D materials \cite{b7,b8}. Effective gas sensors are evaluated based on four key performance metrics: sensitivity, selectivity, response time, and operational stability. Sensitivity refers to the change in the measurable signal upon exposure to a target analyte relative to a baseline, while selectivity denotes the sensor's ability to discriminate between the target gas and other ambient species \cite{b9}--\cite{b11}. These macroscopic attributes are fundamentally governed by microscopic phenomena such as adsorption attributes, lattice distortions, and modifications in electronic properties like the work function and density of states.\\
\indent Motivated by recent theoretical predictions of stable layered In\textsubscript{2}O\textsubscript{3} nanosheets \cite{in2o3}, this work focuses on the gas sensing performance of novel In\textsubscript{2}O\textsubscript{3} monolayer, schematically depicted in Fig.~\ref{schematic_monolayer}. Prior first-principles studies have established that this monolayer is a wide  indirect bandgap semiconductor, with computed bandgaps of 1.64 eV (PBE) and 2.93 eV (HSE06) \cite{in2o3}. Interestingly, although intrinsically nonmagnetic, monolayer In\textsubscript{2}O\textsubscript{3} has been predicted to exhibit ferromagnetism and half-metallicity upon hole doping, a behavior arising from a Van Hove singularity near the valence band edge that leads to a Stoner instability. Monte Carlo simulations further estimate a Curie temperature of up to 62 K under moderate doping, underscoring the promise of In\textsubscript{2}O\textsubscript{3} for future 2D spintronic and multifunctional sensing applications. \\
\indent In this study, we present a comprehensive first-principles investigation of the gas sensing capabilities of the novel In\textsubscript{2}O\textsubscript{3} monolayer, with particular emphasis on its adsorption characteristics and detection potential toward ten hazardous inorganic gases—NH\textsubscript{3}, NO, NO\textsubscript{2}, SO\textsubscript{2}, CS\textsubscript{2,} H\textsubscript{2}S, HCN, CCl\textsubscript{2}O, CH\textsubscript{2}O, and CO. To evaluate the monolayer's performance under ambient conditions, interactions with common background gases such as O\textsubscript{2}, CO\textsubscript{2}, and H\textsubscript{2}O are also examined. Due to the well-established chemical inertness of N\textsubscript{2}, its adsorption behavior is not considered in detail. The analysis aims to assess the viability of the In\textsubscript{2}O\textsubscript{3} monolayer for application in both resistive-type and work function-based gas sensing platforms. \\
\indent The remainder of the paper is organized as follows. Section \ref{method} describes the computational framework and parameters employed for computational analysis. Section \ref{result} presents the core findings on gas adsorption, with subsections \ref{adsorption}, \ref{resistive}, and \ref{workfunction} addressing adsorption properties, resistive sensing response, and workfunction-based sensing, respectively. Section \ref{strain} explores how mechanical strain can be employed to enhance sensing capabilities. A summary of our findings is presented in the conclusion in Section \ref{conclusion}.
\section{Computational Methods} \label{method}
\indent We have used the plane-wave-based Quantum Espresso suite for the simulations. Projector augmented wave (PAW) method,  employing the generalized gradient approximations (GGA)  in the Perdew-Burke-Ernzerhof parameterization (PBE),  was employed for DFT calculations.   We have used a  wavefunction cutoff of $60$Ry. Grimme’s DFT-D3 approach was used to describe non-local van der Waals interactions between the molecule-monolayer system \cite{van1,van2}.  A $3 \times 3 \times 1$ super-cell of In$_2$O$_3$ in the xy plane is employed for investigation on the gas adsorption properties. To avoid spurious
interactions between periodic images and correctly model the gas adsorption properties, a vacuum-pad  of $25$\AA ~~~was employed in the z-direction.  For iterative solution of the Kohn-Sham equations, the energy convergence
threshold was set to be $10^{-6}$Ry, and the forces on all atoms were converged till $10^{-4}$Ry.Bohr$^{-1}$. A $10 \times 10 \times 1$ Monkhorst-grid of k-points was used to sample the Brillouin zone for structural optimization of the unit cell. The k-mesh was adapted accordingly for the supercell. For self-consistent field (SCF) and density of states calculations, a $25 \times 25 \times 1$  Monkhorst-grid of k-mesh were used for the supercell. For work-function calculations, appropriate dipole corrections was employed to compute the local electrostatic potential along the z-direction.\\
\section{Results and Discussion} \label{result}
For effective electrical gas sensing, a target molecule must establish a stable adsorption configuration with the sensing material, which generally requires an adsorption energy  less than $-15k_{B}T$, measuring around $-0.4$eV at room temperature~\cite{b22,afreen}. Furthermore, the interaction between the gas molecule and the substrate should induce a detectable variation in an electrical property of the material. In this section, we present a predictive evaluation of the sensing capabilities of (i) resistive-type and (ii) workfunction-type gas sensors based on the recently proposed In\textsubscript{2}O\textsubscript{3} monolayer \cite{in2o3}.  For effective detection of the target gas  in resistive-type sensors, the adsorption of the gas molecules should lead to the formation of shallow donor or acceptor states that effectively modulate the charge carrier concentration, thereby altering the electrical conductivity of the system~\cite{b6},~\cite{b23}. In contrast, workfunction-type sensors operate on the principle of a change in the surface workfunction upon gas adsorption, which can be measured using suitable apparatus, as discussed in section~\ref{workfunction}. The following analysis of resistive- and workfunction-type gas sensors based on In\textsubscript{2}O\textsubscript{3} will focus on adsorption energetics, density of states (DOS), and surface potential modulation to determine the viability and response characteristics of the material under various gas exposures.
\subsection {Adsorption Characteristics} \label{adsorption}
\begin{figure}
	\centering
	\includegraphics[width=0.8\textwidth]{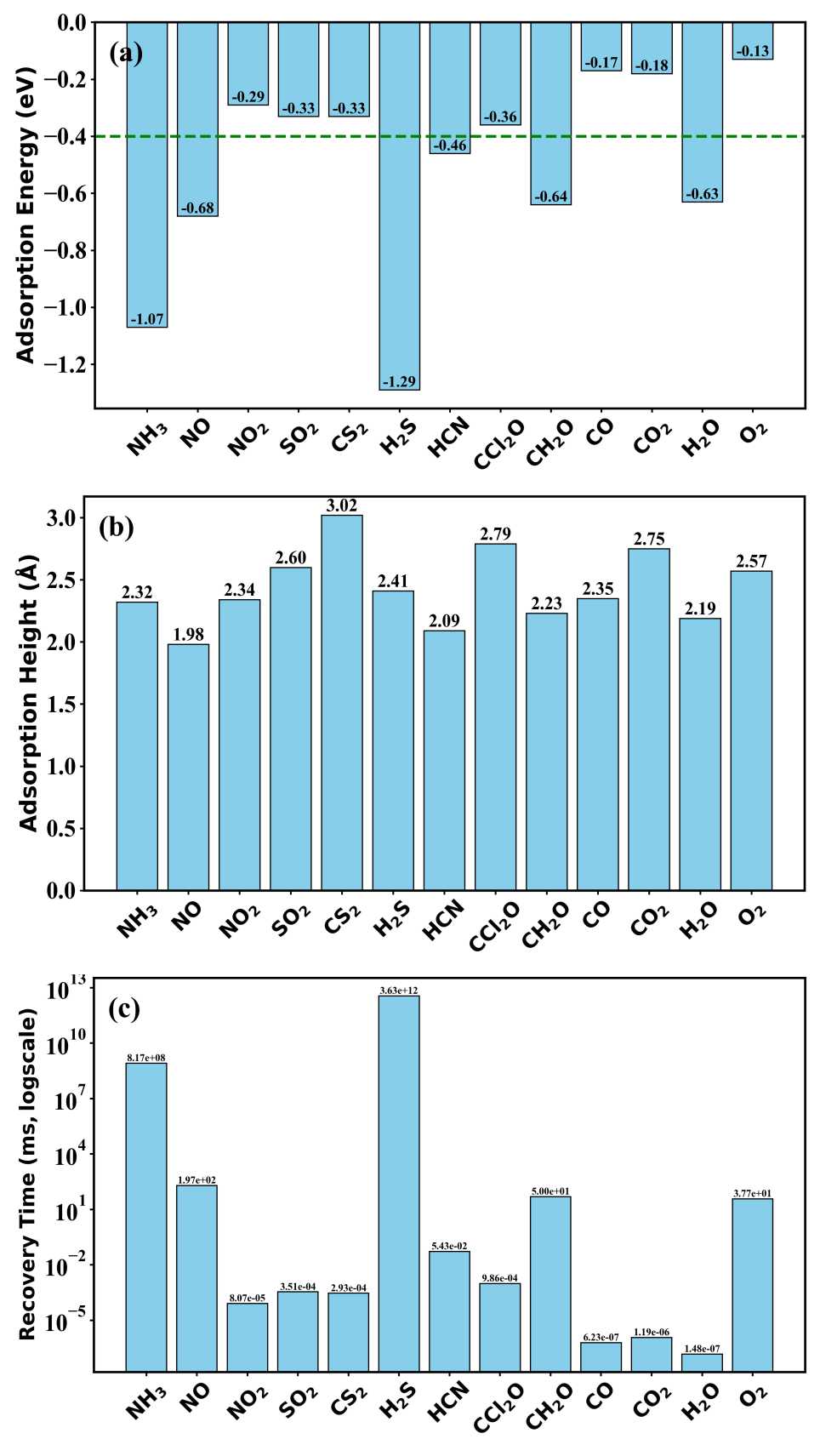}
	\caption{Adsorption characteristics for the considered molecules on In\textsubscript{2}O\textsubscript{3} monolayer. Bar graphs demonstrate (a) adsorption energy (eV), (b) adsorption height (Å), and (c) recovery time for all the gas molecules under consideration.}
	\label{Eads_pic}
\end{figure}

As previously discussed, in both resistive and workfunction-based gas sensing mechanisms, the interaction between gas molecules and the sensing surface requires the formation of a stable adsorbate–adsorbent complex with the underlying material. The adsorption energy serves as a crucial descriptor for evaluating the thermodynamic stability of adsorbed gas species, and it also impacts the  adsorption height. Weak van der Waals forces typically drive physisorption, which is characterized by relatively low adsorption energy values. In contrast, chemisorption involves the formation of stronger chemical bonds, leading to significantly higher adsorption energy magnitudes. In addition to van der Walls forces and chemical bonds, induction of dipole in the monolayer by polar gas molecules, can also influence the adsorption energy. In accordance with established literature, this study adopts a threshold adsorption energy of $-0.4$eV as a criterion to determine whether a gas molecule is thermodynamically stably adsorbed on the surface of the monolayer  \cite{b22, b29, b32}. Adsorption energies under this threshold generally indicate unstable adsorption and recovery time which is too feeble for electrical detection. On the other hand, an adsorption energy exceeding  $-1$eV  generally hints towards the necessity of thermal or UV treatment for recovery of the sensor layer after molecule adsorption  \cite{b22, b29, b32}. Furthermore, an adsorption energy exceeding $-1.5$eV will generally indicate non-reusable or single-use gas sensor or gas scavengers (used for trapping gas molecules or purifying environment). Hence, for our discussions, we will assume a suitable adsorption energy range of $-0.4$eV to $-1$eV for reusable gas sensors.  The adsorption energy ($E_{ads}$) was computed using the following relation:
\begin{equation}
	E_{ads}= E_{molecule+PL} - E_{molecule} - E_{PL},
\end{equation}
where $E_{molecule+PL}$ is the total energy of the system comprising of the gas molecule adsorbed on the parent layer (PL), $E_{molecule}$ is the energy of the isolated gas molecule, and $E_{PL}$ is the energy of the pristine monolayer. Beyond adsorption energy, two additional metrics — adsorption height and recovery time — provide further insight into the interaction dynamics. The adsorption height represents the vertical distance between the adsorbed molecule and the parent layer in the most stable configuration.
To evaluate the reusability of the gas sensors, the recovery time ($\tau$) was estimated via the Arrhenius-type equation:
\begin{equation}
	\tau = \frac{1}{\nu_0 \exp(-E_{ads}/k_B T)},
\end{equation}
where $\nu_0$ denotes the attempt frequency (assumed as $10^{12}$ Hz), $k_B$ is the Boltzmann constant, and $T$ is the operating temperature in Kelvin. For practical sensor applications, a recovery time on the order of tens to hundreds of milli-seconds is generally preferred \cite{b29},\cite{b34}. Two-dimensional materials offer multiple potential sites for molecular adsorption. However, the energetically most favorable configuration — i.e., the one exhibiting the maximum magnitude of adsorption energy — is typically considered the most stable. In this study, the favourable adsorption site for each gas molecule was systematically explored by initially placing the molecule at multiple strategic orientations (both horizontal and vertical) on the high-symmetry positions of the In\textsubscript{2}O\textsubscript{3} monolayer, as demonstrated in Fig.~\ref{schematic_monolayer} (and   Section A1 of the Supplementary Material). The most stable configuration, corresponding to the minimum adsorption energy, was identified and adopted for subsequent analysis (details provided in Section A1 of Supplementary Material). The top and side views of the monolayer with adsorbed molecules in the most stable (minimum energy) configuration are demonstrated in Section A2 of the Supplementary material. The optimized adsorption energy, adsorption height and recovery time for each gas molecule on the In\textsubscript{2}O\textsubscript{3} monolayer is summarized in Table~\ref{ads_table} and demonstrated graphically in Fig.~\ref{Eads_pic}. \\
\indent All thirteen molecules examined in this study exhibit negative adsorption energies in their most energetically favourable configurations. This observation confirms   the exothermic nature of the adsorption process. Based on the calculated adsorption energies given in Table \ref{ads_table}, it is evident that NO\textsubscript{2}, SO\textsubscript{2}, CS\textsubscript{2}, CCl\textsubscript{2}O, CO, CO\textsubscript{2}, and O\textsubscript{2} exhibit weak interactions with the monolayer, as their adsorption energies lie below the widely acknowledged  threshold of $-0.4~{eV}$, suggesting that these molecules may not be effectively retained on the surface for reliable detection. In particular,  CO, CO\textsubscript{2}, and O\textsubscript{2}, demonstrate adsorption energies of $-0.17~\text{eV}$, $-0.18~\text{eV}$, and $-0.13~\text{eV}$, along with adsorption heights of $2.35~\text{\AA}$, $2.75~\text{\AA}$, and $2.57~\text{\AA}$, respectively. NO\textsubscript{2}, SO\textsubscript{2},  CS\textsubscript{2} and CCl\textsubscript{2}O exhibited slightly higher adsorption energies of $-0.29~\text{eV}$, $-0.33~\text{eV}$,  $-0.33~\text{eV}$ and $-0.36$~eV, respectively, still remaining under the threshold adsorption energy  of $-0.4$eV. The corresponding adsorption heights for NO\textsubscript{2}, SO\textsubscript{2},  CS\textsubscript{2} and CCl\textsubscript{2}O are found to be $2.34~\text{\AA}$, $2.60~\text{\AA}$,  $3.02~\text{\AA}$, and $2.79~\text{\AA}$ respectively. It is important to note that recovery times for these molecules, of the order of one to tens of microseconds, are generally too brief to produce measurable  changes in electrical conductivity or workfunction in commercial set-ups, thereby limiting the effectiveness of reliable gas detection in such cases. \\
\indent HCN, with an adsorption energy of $-0.46~\text{eV}$ and a height of $2.09~\text{\AA}$, demonstrates a relatively moderate interaction with the monolayer. Molecules such as CH\textsubscript{2}O and H\textsubscript{2}O exhibit adsorption energies of $-0.64~\text{eV}$ and $-0.63~\text{eV}$, respectively, with moderate recovery times of 50 ms and 37.7 ms and adsorption heights around 2.2~\text{\AA}, indicating that both gases are moderately bound yet can be readily desorbed, making them potential candidates for reusable room-temperature gas sensing with In\textsubscript{2}O\textsubscript{3} monolayer. Among all molecules, NO shows the shortest adsorption height of $1.98~\text{\AA}$ and a moderate adsorption energy of $-0.68~\text{eV}$, yet it induces minimal distortion in the monolayer structure. Its recovery time of 197~ms further supports its suitability for real-time sensing, striking an optimal balance between stability, and reusability. \\ 
\begin{figure*}[!htb]
	\centering
	\includegraphics[width=\textwidth]{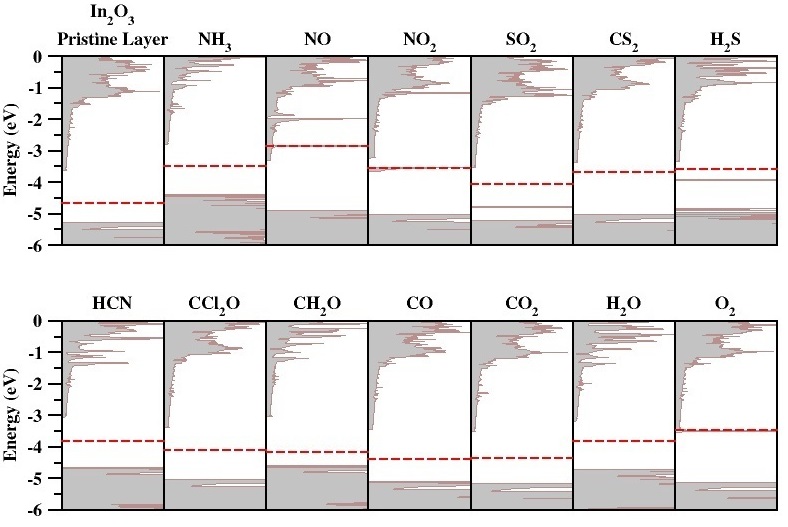}
	\caption{Total density of states (gray shaded region) for the pristine  In\textsubscript{2}O\textsubscript{3} monolayer and the combined molecule-monolayer systems.  The  Fermi level has been shown by red dashed lines.  }
	\label{dos_all}
\end{figure*}
\indent The two molecules, NH\textsubscript{3} and H\textsubscript{2}S, exhibit very high adsorption energies of $-1.07~\text{eV}$ and $-1.29~\text{eV}$, respectively, and induce notable structural distortion, consistent with their higher adsorption energy. The adsorption heights of these molecules are  2.32\AA~ and 2.41{\AA} respectively. Despite strong binding, their long recovery times of $8.17\times 10^8 ms$ and $3.63\times 10^{12}$ms pose significant limitations for practical reuse as reusable sensors without thermal or UV treatment. It is noteworthy that the ambient molecule H\textsubscript{2}O exhibits a moderate adsorption energy of $-0.63$eV, exceeding the  threshold of $-0.4$eV. This suggests a potential for competitive adsorption of H\textsubscript{2}O at the active sites, which may hinder the selective detection of target toxic gases. Therefore, the In\textsubscript{2}O\textsubscript{3} monolayer sensor may not be suitable for deployment in humid environments. \\
\begin{table}[!htb]
	\begin{center}
		\caption{Values of Adsorption Energy ($E_{ads}$), Adsorption Height (H) and Recovery time ($\tau$),   for different molecules investigated on parent $In\textsubscript{2}O\textsubscript{3}$ layer.}
		\begin{tabular}{ |c|c|c|c| } 
			\hline
			\textbf{Molecule} & \textbf{E\textsubscript{ads}~(eV)}  & \textbf{H(\AA)} & \textbf{$\tau$ (ms)}  \\
			\hline
			NH\textsubscript{3} & -1.07 & 2.32  & $8.17\times$$10^{8}$  \\ 
			\hline
			NO & -0.68 & 1.98  & 197  \\ 
			\hline
			NO\textsubscript{2} & -0.29 & 2.34 & $8.07\times$$10^{-5}$   \\ 
			\hline
			SO\textsubscript{2} & -0.33 & 2.60  & $3.51\times$$10^{-4}$  \\ 
			\hline
			CS\textsubscript{2} &  -0.33 & 3.02 & $2.93\times$$10^{-4}$  \\ 
			\hline
			H\textsubscript{2}S & -1.29 & 2.41 & $3.63\times$$10^{12}$  \\ 
			\hline
			HCN &  -0.46 & 2.09 & $5.43\times$$10^{-2}$  \\ 
			\hline
			CCl\textsubscript{2}O &  -0.36 & 2.79 &  $9.86\times$$10^{-4}$ \\ 
			\hline
			CH\textsubscript{2}O &  -0.64 & 2.23 &  50.0 \\ 
			\hline
			CO & -0.17 & 2.35 & $6.23\times$$10^{-7}$  \\ 
			\hline
			CO\textsubscript{2} &  -0.18 & 2.75  & $1.19\times$$10^{-6}$  \\ 
			\hline
			H\textsubscript{2}O &  -0.63 & 2.19  & 37.7  \\ 
			\hline
			O\textsubscript{2} &  -0.13 & 2.57 & $1.48\times$$10^{-7}$  \\ 
			\hline
			
		\end{tabular}
		\label{ads_table}
	\end{center}
\end{table}
\indent For comparison,  comprehensive tables of the adsorption energy and  recovery times for the potential molecules, demonstrating adsorption energy beyond $-0.4$eV (namely, NH\textsubscript{3}, NO, H\textsubscript{2}S, HCN, CH\textsubscript{2}O and H\textsubscript{2}O) on various 2D materials (as reported in the literature), has been included in Sec. A3 of the Supplementary Material. 
\subsection {Application in resistive-type gas sensing} \label{resistive}
A resistive-type gas sensor operates by detecting changes in electrical resistance caused by the adsorption of gas molecules onto its surface. This process of change in resistance may follow the formation of shallow donor or acceptor states in the band-gap, promoting additional electrons or holes into the conduction or valence bands respectively, and thereby increasing the concentration of mobile charge carriers that modulate the system’s resistance. To gain insight into the sensing mechanism of the In\textsubscript{2}O\textsubscript{3} monolayer, we analyze the corresponding changes in the density of states (DOS) upon gas adsorption. Figure~\ref{dos_all} shows the total DOS (in gray), with the Fermi level denoted by a red dashed line. The resulting modifications in the DOS and relative shift in Fermi-energy offer a qualitative understanding of the conductivity variations observed upon gas molecule interaction, as these directly influence carrier availability and transport properties.\\
\indent Figure~\ref{dos_all} reveals that NO,  and O\textsubscript{2} adsorption induces metallic behavior in the 2D In\textsubscript{2}O\textsubscript{3} monolayer. For NO and O\textsubscript{2}, the Fermi level penetrates inside the conduction band. These shifts suggest a substantial increase in electrical conductivity upon adsorption of the gas molecules.  Among these, NO shows the most favorable sensing performance, with an adsorption energy of -0.68eV making it highly effective for room-temperature detection. In the case of O\textsubscript{2}, the weak interaction, reflected by an adsorption energy of -0.13~eV and rapid recovery time, limits its effectiveness for sensing under ambient conditions. \\
\indent A pronounced change in conductivity is also expected upon NO\textsubscript{2} adsorption, as the formation of a partially filled state in close proximity to the conduction band edge reduces the bandgap to 0.35~eV. This significant bandgap narrowing lowers the energy barrier for electron excitation into the conduction band, thereby promoting enhanced carrier generation and contributing to improved sensing response.  However,  the weak adsorption interaction, reflected by an adsorption energy of -0.29~eV and fast recovery time, restricts its suitability for reliable sensing under ambient conditions.  For H\textsubscript{2}S and SO\textsubscript{2} adsorption, localized filled states emerge within the band gap. Particularly, in the case of H\textsubscript{2}S, an induced state is located near the middle of the bandgap, slightly shifted toward the conduction band, whereas for SO\textsubscript{2}, the induced state appears close to the valence band maximum. The donor state generated by H\textsubscript{2}S, hence,  effectively narrows the bandgap, introducing a moderately shallow defect level that facilitates electron excitation into the conduction band. However, the  state induced by SO\textsubscript{2} in the bandgap is a deep donor level and fails to  significantly enhance conductivity at room temperature.  H\textsubscript{2}S  adsorption is also characterized by a  high adsorption energy of -1.29eV on the 2D In\textsubscript{2}O\textsubscript{3} monolayer. Thus the monolayer can detect  H\textsubscript{2}S, but the high adsorption energy may hamper the reusability of the sensor unless subjected to thermal treatment or UV radiation for recovery. The adsorption of SO\textsubscript{2} is relatively weak, with an adsorption energy of -0.33~eV, rendering its detection by the monolayer ineffective at room temperature.\\
In the cases of NH\textsubscript{3}, H\textsubscript{2}O, HCN, CO, CO\textsubscript{2}, CCl\textsubscript{2}O, CS\textsubscript{2}, and CH\textsubscript{2}O molecules, adsorption on the 2D In\textsubscript{2}O\textsubscript{3} monolayer does not induce any shallow donor or acceptor states, nor does it cause significant bandgap narrowing relative to the pristine system. As a result, the electronic structure remains largely unperturbed, and no substantial change in conductivity is expected upon adsorption of these molecules. Although NH\textsubscript{3},  CH\textsubscript{2}O and H\textsubscript{2}O exhibit a good adsorption energy of -1.07eV, -0.64eV and  -0.63eV respectively,  none of these molecules lead to the formation of electronically active states capable of enhancing charge carrier concentration at room temperature. \\
\indent These results highlight the 2D In\textsubscript{2}O\textsubscript{3} monolayer as a promising platform for resistive-type gas sensing, with pronounced sensitivity toward NO and H\textsubscript{2}S driven by adsorption-induced electronic modulation, while its limited response to other analytes ensures intrinsic selectivity.\\
\indent \textbf{Conductivity Change Factor for the Adsorbed Systems:}
In several cases discussed above, the formation of induced empty or filled states within the bandgap of the pristine monolayer can facilitate enhanced conductivity. These states lower the required energy for electronic transitions from the highest occupied molecular orbital (HOMO) to the lowest unoccupied molecular orbital (LUMO), thereby enabling easier excitation of electrons into the conduction band or holes into the valence band. To quantitatively evaluate the electrical response of gas adsorption within the framework of resistive-type sensing, we define the conductivity change factor, $\chi$, as the ratio of the electrical conductivity of the In\textsubscript{2}O\textsubscript{3} monolayer with the adsorbed gas molecule to that of its pristine form. The electrical conductivity can be approximated using the expression:
\begin{equation}
	\sigma = A T^{3/2} e^{-E_{g}/2k_{B}T},
\end{equation}
\indent where \textit{A} is a material-specific constant, \textit{T} represents the absolute temperature, $k_B$ is the Boltzmann constant, and $E_g$ denotes the energy difference between the highest occupied molecular orbital (HOMO) and the lowest unoccupied molecular orbital (LUMO) for systems exhibiting semiconducting behavior~\cite{b34}. \\
\indent For molecule–monolayer systems exhibiting metallic character due to partially filled localized states within the bandgap, $E_g$ is instead defined as the smaller of the two energy separations: between the induced state and the valence band maximum, or between the induced state and the conduction band minimum. Since the conductivity of a material depends on the density of charge carriers (electrons in the conduction band and holes in the valence band), which are thermally activated across $E_g$, any reduction in this gap upon gas adsorption enhances carrier excitation and thus increases conductivity.
The corresponding conductivity change factor $\chi$ can be expressed as: 
\begin{equation}\label{eq:chi}
	\chi = \exp \left[ \frac{-(E_g - E_g')}{2k_B T} \right]
\end{equation}
where $E_g'$ is the bandgap of the pristine In\textsubscript{2}O\textsubscript{3} monolayer, and $E_g$ corresponds to the modified bandgap upon gas adsorption as previously defined. The computed $E_g$ and $\chi$ values for all adsorbed systems are summarized in Table~\ref{Eg_table}. Since the intrinsic conductivity of the In\textsubscript{2}O\textsubscript{3} monolayer is inherently low, a conductivity change factor ($\chi$) exceeding $10^6$ is necessary to enable effective detection in cost-efficient commercial devices.\\
\indent For NO and O\textsubscript{2} adsorption, Eq.~\eqref{eq:chi} becomes inapplicable as the Fermi level shifts into the conduction band, inducing metallic behavior with intrinsically high conductivity. The NO\textsubscript{2}-adsorbed system demonstrates an exceptional conductivity modulation, with a change factor ($\chi$) of $ 2.13 \times 10^{11}$. In addition, for H\textsubscript{2}O adsorbed system $\chi$ takes the value of $7.86 \times 10^8$, which when combined with its high adsorption energy translates into a  strong detection potential on the monolayer. In contrast to the above molecules, SO\textsubscript{2} induces a relatively modest conductivity change of $\sim 2.76 \times 10^3$, which, given the inherently low conductivity of the pristine monolayer, may be insufficient for reliable detection in cost-effective commercial sensors.  For the remaining analytes, the calculated $\chi$ values are are very low (less than 10), rendering them unsuitable for resistive-type detection based on conductivity variations alone.  As elaborated in Section~\ref{adsorption}, despite their strong conductivity response, the low adsorption energies of NO\textsubscript{2} and O\textsubscript{2} compromise their practical detectability via resistive-type sensing, due to poor surface retention and rapid desorption. Consequently, among all the studied gases, NO and H\textsubscript{2}S emerge as the most promising candidates for room-temperature resistive-type gas sensing, owing to their favorable combination of significant DOS modifications, large conductivity change factors, and adequate adsorption strengths. 
\begin{table}[h]
	\begin{center}
		\caption{Values of $E_g$ and $\chi$ for different molecules investigated on parent layer  In\textsubscript{2}O\textsubscript{3}.}
		\begin{tabular}{ |c|c|c| } 
			\hline
			\textbf{Molecule} & \textbf{E\textsubscript{g} (eV)} & \textbf{$\chi$}   \\
			
			\hline
			Pristine In\textsubscript{2}O\textsubscript{3} &   1.65 & 1.0   \\ 
			\hline
			NH\textsubscript{3} & 1.59 & 3.19   \\ 
			\hline
			NO & 0 &  very high  \\ 
			\hline
			NO\textsubscript{2} & 0.3 &  2.13$\times$$10^{11}$  \\ 
			\hline
			SO\textsubscript{2} & 1.24 & 2.76$\times$$10^{3}$   \\ 
			\hline
			CS\textsubscript{2} &  1.66  & 0.82 \\ 
			\hline
			H\textsubscript{2}S & 0.59 & 7.86$\times$$10^{8}$   \\ 
			\hline
			HCN & 1.60 & 2.63 \\ 
			\hline
			CCl\textsubscript{2}O &  1.66 & $0.82$  \\ 
			\hline
			CH\textsubscript{2}O &  1.57 & 4.69  \\ 
			\hline
			CO &  1.65 & 1.0  \\ 
			\hline
			CO\textsubscript{2} &  1.65 & 1.0  \\ 
			\hline
			H\textsubscript{2}O &  1.55 & 6.91  \\ 
			\hline
			O\textsubscript{2} &  0 & very high  \\ 
			\hline
		\end{tabular}
		\label{Eg_table}
	\end{center}
\end{table}
\subsection {Application in Work Function Type Gas Sensing} \label{workfunction}
\begin{figure}
	\centering
	\includegraphics[width=0.8\textwidth]{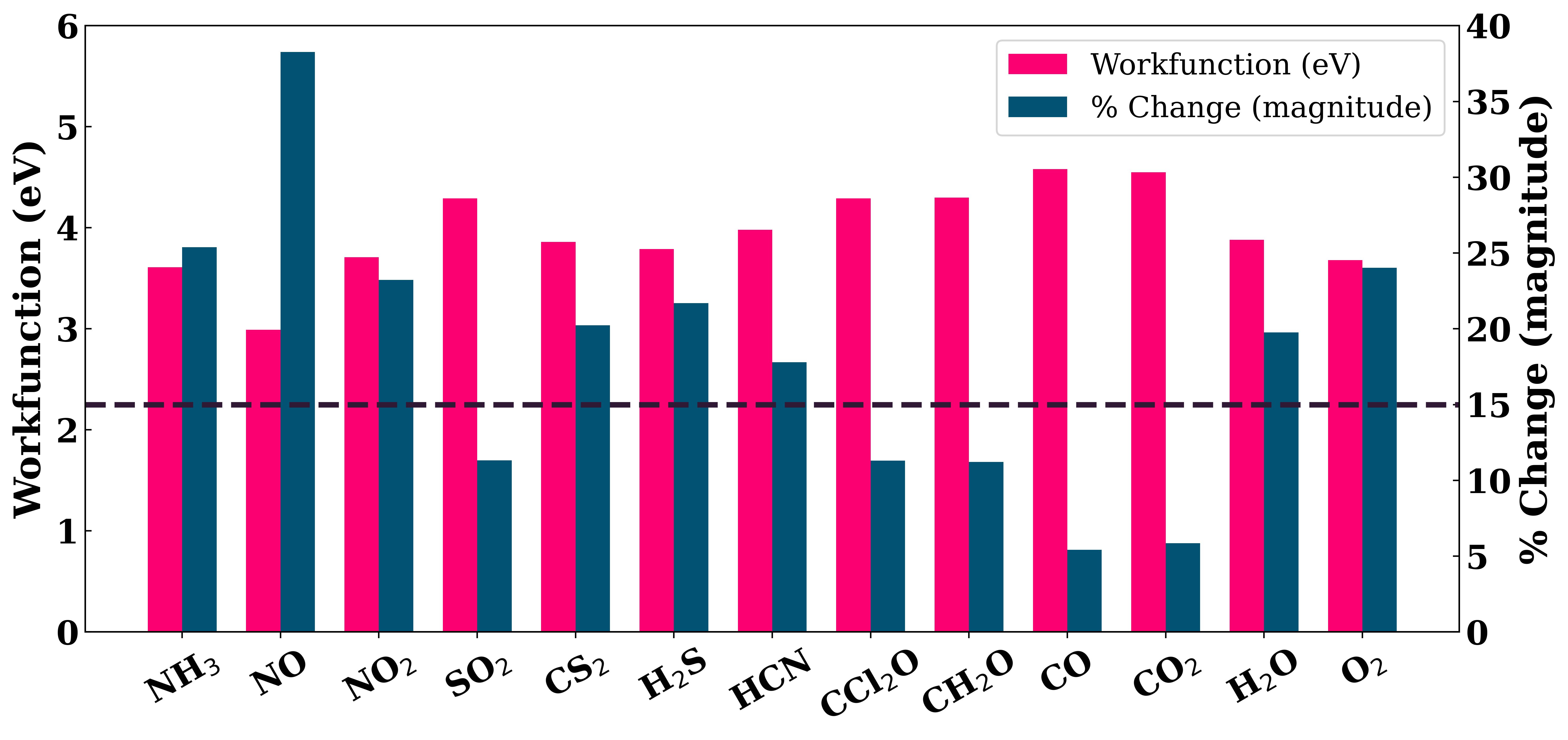}
	\caption{Percentage change in the workfunction of the In$_2$O$_3$ monolayer upon adsorption of gas molecules, compared to the pristine monolayer. The magenta dashed line represents the minimum percentage change in workfunction required for efficient detection.}
	\label{wf_molecules}
\end{figure}

A critical parameter for evaluating the sensing capabilities of the 2D-In$_2$O$_3$ monolayer is the change in work function ($\Phi$) induced by gas adsorption. This metric reflects the shift in the surface electronic potential upon interaction with gas molecules and serves as an indicator of charge redistribution and dipole formation at the interface. The principle of this sensing mechanism is based on the Kelvin probe technique, where variations in the contact potential difference are used to measure the work function with high precision.\\
\indent The work function is calculated as the difference between the vacuum potential and the Fermi level of the system, as shown in the equation:
\begin{equation}
	\phi = E_{vac} - E_{f}
\end{equation}
where $\phi$ is the work function, $E_{vac}$ is the vacuum energy level, and $E_{f}$ is the Fermi energy. For the pristine In$_2$O$_3$ monolayer, the computed work function is 4.84 eV. The change in work function between the pristine monolayer and the adsorbed system quantifies the sensor's response, with larger shifts indicating higher sensitivity. In our work, a threshold of 15\% workfunction change on the adsorption of the gas molecule is considered the minimum detectable change. This is because smaller shifts are typically insufficient for reliable detection with conventional techniques. \\
\indent Fig.~\ref{wf_molecules} and Table~\ref{Wf_molecules_table} demonstrate the absolute workfunction of the gas adsorbed monolayer as well as the percentage change in workfunction with respect to the pristine layer. It can be noted that the adsorption of NO induces the highest workfunction change of 38.27\%, with a calculated value of 2.99 eV, emphasizing the high sensitivity of the 2D-In$_2$O$_3$ monolayer for NO detection. Additionally, adsorption of NH$_3$, NO$_2$, CS$_2$, H$_2$S, HCN, H$_2$O, and O$_2$ results in workfunction changes exceeding 15\%, with values of 25.38\%, 23.23\%, 20.24\%, 21.7\%, 17.8\%, 19.77\%, and 24.02\%, respectively. Out of these molecules, NH$_3$, NO, H$_2$S, HCN, and H$_2$O have adsorption energy in excess of $-0.4eV$, based on the adsorption energy profile reported in Sec.~\ref{adsorption}. Hence, out of the toxic gas  molecules,  NH$_3$, NO, H$_2$S, HCN, can be detected using the workfunction based  sensing mechanism. Notably, NH$_3$ and HCN, which remained undetectable using resistance-based methods, now exhibits measurable detection sensitivity through the workfunction-based approach, owing to its high adsorption energy. In addition, since H\textsubscript{2}O posses moderate adsorption energy and demonstrates suitable workfunction change ($>15\%$), the monolayer shouldn't be used as a toxic gas sensor in humid environment, unless the focus is towards detection of H\textsubscript{2}O.
\subsection{Effect of Mechanical Strain on Sensing Performance} \label{strain}
To explore the possibility of further enhancing the gas sensing performance in In$_2$O$_3$ monolayer,  we explore the effect of mechanical strain on the sensor performance.  Strain-driven modulation of  sensitivity is a widely recognized approach in 2D materials research, capable of substantially altering adsorption energetics and electronic structure. In this study, an in-plane biaxial tensile and compressive strain (upto $\pm5\%$) was applied to the monolayer to explore its effect on the sensing performance, particularly focusing on changes in adsorption energy and sensing characteristics. In what follows, we discuss the impact of strain on the adsorption energies of the gas molecules under consideration on the monolayer, and how strain influences their detectability in both resistive-type and work-function-based gas sensing mechanisms.
\subsubsection*{Strain-Induced Modulation of the Adsorption energy} 
\begin{figure}
	\centering
	\includegraphics[width=0.8\textwidth]{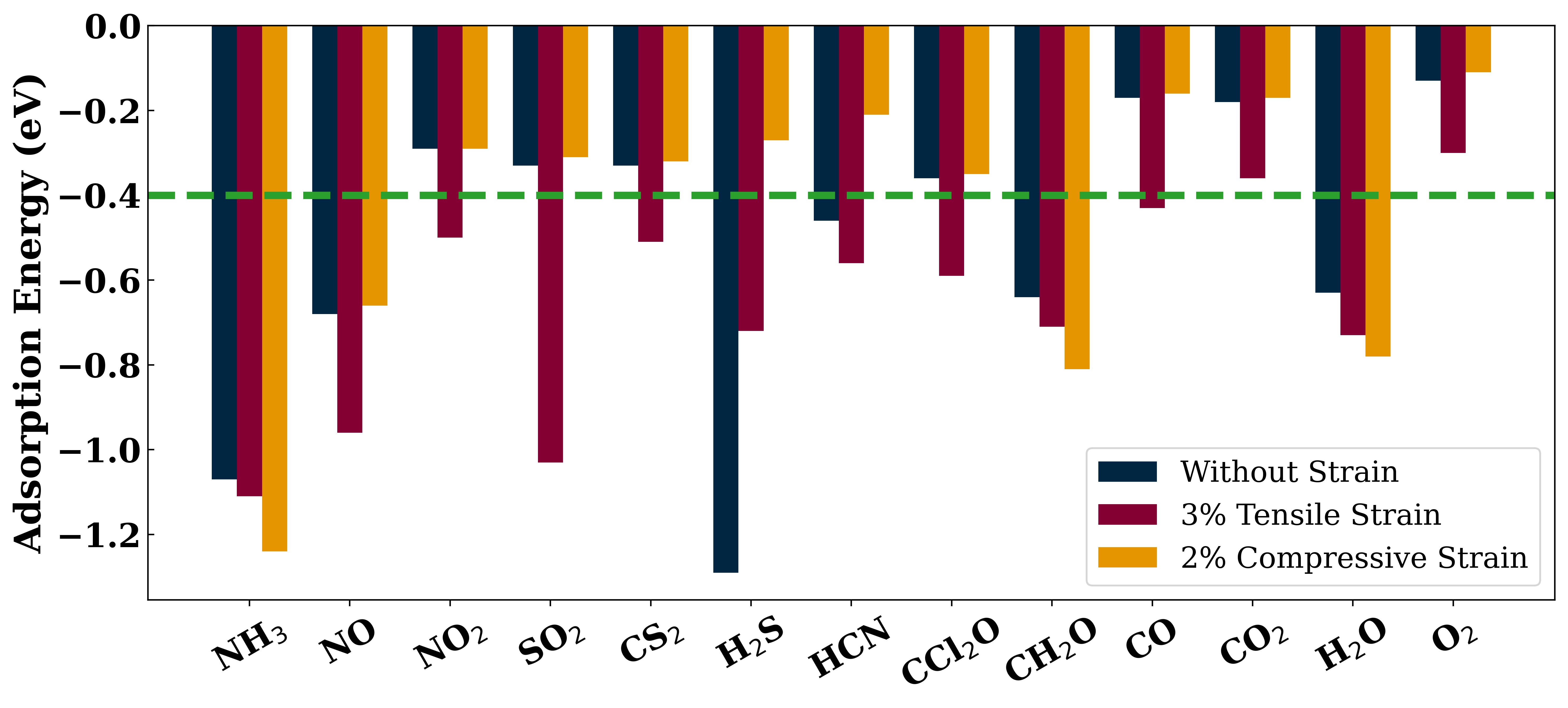}
	\caption{Adsorption energies of gas molecules on the In$_2$O$_3$ monolayer for three cases: unstrained (blue), with 3\% tensile strain (maroon), and with 2\% compressive strain (orange). The green dashed line represents the threshold value of 10$k_B T$ (eV).}
	\label{fig_strain_ads_energy}
\end{figure}

It was observed that with  a tensile strain of approximately 4\%, the adsorption energies of O$_2$ and CO\textsubscript{2} were modulated to $-0.47$eV and $-0.54$eV, both exceeding the normally accepted threshold adsorption energy of $-0.4$eV for suitable electrical detection. Given that O$_2$ is an abundant ambient gas, and that oxygen-adsorbed In\textsubscript{2}O\textsubscript{3} exhibits metallic behavior according to its density of states profile, such moderate adsorption may compromise the sensor’s selectivity by enabling unintended detection of background ambient gases. Moreover, with such moderate adsorption energy, the persistent presence of O$_2$ at the adsorption sites could obstruct the binding of more relevant target analytes, thereby deteriorating the sensor's efficiency and overall detection accuracy. Hence, we conclude that a moderate tensile strain  of upto 3\% is appropriate for investing the potential towards enhanced sensing of the harmful gases under consideration. Similarly, a compressive strain of $3\%$ or more enhances the adsorption energy of H\textsubscript{2}O beyond $-1$eV, which may result in H\textsubscript{2}O molecules to cling for excessively long duration to the active adsorption sites in the monolayer and hamper the detection of harmful gases. Thus, we conclude that a compressive strain of upto 2\% can be suitably applied to the monolayer. \\
\indent  Fig \ref{fig_strain_ads_energy} demonstrates the changes in adsorption energy for the thirteen gas molecules under consideration with 3\% tensile strain and 2\% compressive strain applied to the monolayer. The exact values of the adsorption energy on the monolayer with 3\% tensile strain and 2\% compressive strain is documented in the Table~\ref{tab:adsorption_energy}. It is evident that the application of 3\% tensile strain significantly improves the adsorption behavior of several gas molecules on the In$_2$O$_3$ monolayer. In addition to molecules  such as NH$_3$, NO, H\textsubscript{2}S, HCN, CH\textsubscript{2}O and H\textsubscript{2}O, that already exhibited  adsorption energies exceeding $-0.4$eV in the unstrained monolayer, molecules such as  NO$_2$, SO$_2$, CS$_2$, CCl$_2$O, and CO, achieved adsorption energies exceeding $-0.4$eV  under 3\% tensile strain, indicating stable adsorbate–adsorbent configuration for suitable detection. Among these molecules, SO$_2$ stands out, with its adsorption energy increasing sharply to $-1.03$~eV. The other molecules NO$_2$, CS$_2$, CCl$_2$O, and CO demonstrate adsorption energy of $-0.5$eV, $-0.51$eV, $-0.59$eV and $-0.43$eV with 3\% tensile strain. Interestingly, H\textsubscript{2}S, which initially exhibited an adsorption energy of $-1.29$eV on the unstrained monolayer, now shows a reduced adsorption energy of $-0.72$eV on the monolayer with 3\% tensile strain, rendering it more suitable for a reusable gas sensor. In addition, the adsorption energy of NO increases from $-0.68$eV to $-0.96$eV under 3\% tensile strain. We also note the undesirable increase in adsorption energy for H\textsubscript{2}O molecule from $-0.63$eV to $-0.73$eV due to the applied tensile strain. \\
\indent With a 2\% compressive strain  applied to the In$_2$O$_3$ monolayer, most gas molecules exhibit same or  a decrease in adsorption energy, with several falling below the threshold value of  $-0.4$eV, except NH$_3$, CH$_2$O, and H$_2$O. These molecules, namely NH$_3$, CH$_2$O, and H$_2$O, exhibit increased adsorption energies  of $-1.24$eV, $-0.81$eV and $-0.78$eV respectively. \\
\subsubsection*{Resistive type gas sensing with applied strain} The density of states of the monolayer without and with the adsorbed gas molecules under 3\% tensile strain and 2\% compressive strain are demonstrated in Fig.~S4 and Fig.~S5 of the Supplementary Meterial.    Among the molecules NO$_2$, SO$_2$, CS$_2$, CCl$_2$O, and CO, that demonstrated  crossing the threshold adsorption energy of $-0.4$eV at 3\% tensile strain, NO\textsubscript{2} adsorption features an induced  filled state near the edge of the conduction band. This induced state, thus, acts as a shallow donor and enhances the conductivity.  The bandgap between the induced state and the conduction band edge was found to be $0.3$~eV, giving rise to a large  conductivity change factor ($\chi$) of almost $2.13\times 10^11$.  CS\textsubscript{2} adsorption demonstrates a filled state in the bandgap of the pristine In\textsubscript{2}O\textsubscript{3}, which modifies the effective bandgap to $0.76$~eV, increasing the conductivity of the monolayer by an approximate factor ($\chi$) of $5\times 10^4$. However, since the monolayer has very low conductivity to begin with, such an enhancement in conductivity might not be suitable for detection with a commercially deployable low cost-set-up. The other molecule, such as, SO\textsubscript{2}, CCl\textsubscript{2}O, and CO  adsorption, on the other hand, doesn't demonstrate suitable features in the DOS profile for electrical detection. \\
\indent On the application of 2\% compressive strain, no new molecule cross the adsorption energy threshold of $-0.4$eV. Hence,  compressive strain does not offer any significant advantage for gas sensing applications with In\textsubscript{2}O\textsubscript{3} monolayer. \\
\indent Thus, we conclude that the application of 3\% tensile strain to the In\textsubscript{2}O\textsubscript{3} monolayer aids the detection towards NO\textsubscript{2}  molecules, which remain undetectable in the unstrained configuration. Additionally, the tensile strain reduces the adsorption energy of the H\textsubscript{2}S molecule to a moderate range, thereby ensuring the sensor's reusability for H\textsubscript{2}S detection. In contrast, the application of compressive strain does not confer any appreciable benefit for gas sensing applications.
\begin{table}
	\begin{center}
		\caption{The computed work function values for all gas-adsorbed configurations, along with their respective percentage deviations from the pristine In\textsubscript{2}O\textsubscript{3} monolayer.}
		\begin{tabular}{ |c|c|c| } 
			\hline
			\textbf{Molecule} & \textbf{Workfunction (eV)}  & \textbf{\% Change} \\
			\hline
			{Pristine In\textsubscript{2}O\textsubscript{3}} & 4.84 & 0 \\
			NH$_3$      & 3.61 & 25.38 \\
			NO          & 2.99 & 38.27 \\
			NO$_2$      & 3.71 & 23.23 \\
			SO$_2$      & 4.29 & 11.33 \\
			CS$_2$      & 3.86 & 20.24 \\
			H$_2$S      & 3.79 &21.70 \\
			HCN         & 3.98 & 17.80 \\
			CCl$_2$O    & 4.29 & 11.30 \\
			CH$_2$O     & 4.30 & 11.21 \\
			CO          & 4.58 & 5.43 \\
			CO$_2$      & 4.55 & 5.86 \\
			H$_2$O      & 3.88 & 19.77 \\
			O$_2$       & 3.68 & 24.02 \\
			\hline
		\end{tabular}
		\label{Wf_molecules_table}
	\end{center}
\end{table}

\begin{figure}[h]
	\centering
	\includegraphics[width=0.8\textwidth]{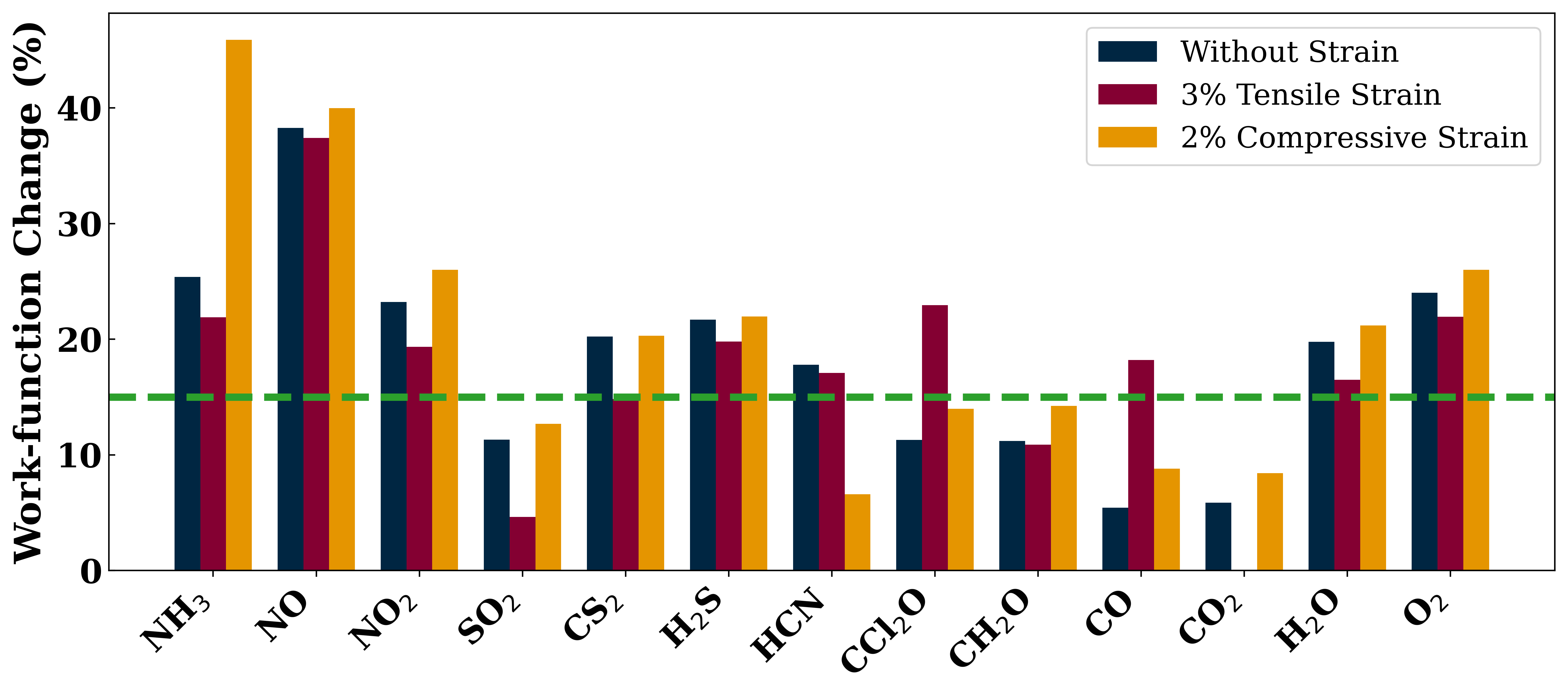}
	\caption{Percentage change in the workfunction of the In$_2$O$_3$ monolayer upon adsorption of gas molecules, compared to the pristine monolayer, for three cases: unstrained (blue), with 3\% tensile strain (maroon), and with 2\% compressive strain (orange). The green dashed line represents the minimum percentage change in workfunction required for efficient detection.}
	\label{fig_strain_wf}
\end{figure}

\begin{table}[h!]
	\centering
	\caption{Adsorption Energies (eV) of Gas Molecules on In$_2$O$_3$ Monolayer under Different Strain Conditions}
	\label{tab:adsorption_energy}
	\begin{tabular}{|c|c|c|c|}
		\hline
		\multirow{2}{*}{\textbf{Molecule}}  & \multicolumn{3}{c|}{Adsorption energy (eV)}\\
		\cline{2-4}
		& \textbf{No Strain} & \textbf{+3\% Strain} & \textbf{-2\% Strain} \\
		\hline
		NH$_3$    & -1.07 & -1.11 & -1.24 \\
		NO        & -0.68 & -0.96 & -0.66 \\
		NO$_2$    & -0.29 & -0.50 & -0.29 \\
		SO$_2$    & -0.33 & -1.03 & -0.31 \\
		CS$_2$    & -0.33 & -0.51 & -0.32 \\
		H$_2$S    & -1.29 & -0.72 & -0.27 \\
		HCN       & -0.46 & -0.56 & -0.21 \\
		CCl$_2$O  & -0.36 & -0.59 & -0.35 \\
		CH$_2$O   & -0.64 & -0.71 & -0.81 \\
		CO        & -0.17 & -0.43 & -0.16 \\
		CO$_2$    & -0.18 & -0.36 & -0.17 \\
		H$_2$O    & -0.63 & -0.73 & -0.78 \\
		O$_2$     & -0.13 & 0.30  & -0.11 \\
		\hline
	\end{tabular}
\end{table}

\begin{table}[h]
	\centering
	\caption{Workfunction values and percentage changes for various molecules under tensile and compressive strain}
	\label{tab:wf_strain}
	\begin{tabular}{|c|c|c|c|c|}
		\hline
		\multirow{2}{*}{Molecule} & \multicolumn{2}{c|}{+3\% Strain} & \multicolumn{2}{c|}{-2\% Strain} \\
		\cline{2-5}
		& W.F. (eV) & \% Change & W.F. (eV) & \% Change \\
		\hline
		In$_2$O$_3$ & 4.76 & - & 4.85 & - \\
		\hline
		NH$_3$ & 3.71 & 21.90 & 2.62 & 45.90 \\
		\hline
		NO & 2.98 & 37.40 & 2.91 & 39.98 \\
		\hline
		NO$_2$ & 3.84 & 19.36 & 3.59 & 26.01 \\
		\hline
		SO$_2$ & 4.54 & 4.65 & 4.23 & 12.69 \\
		\hline
		CS$_2$ & 4.05 & 14.83 & 3.86 & 20.31 \\
		\hline
		H$_2$S & 3.81 & 19.80 & 3.78 & 21.96 \\
		\hline
		HCN & 3.94 & 17.09 & 5.17 & 6.60 \\
		\hline
		CCl$_2$O & 3.67 & 22.94 & 4.17 & 13.98 \\
		\hline
		CH$_2$O & 4.24 & 10.88 & 4.16 & 14.24 \\
		\hline
		CO & 3.89 & 18.21 & 4.42 & 8.81 \\
		\hline
		CO$_2$ & 4.76 & 0.02 & 4.44 & 8.43 \\
		\hline
		H$_2$O & 3.97 & 16.49 & 3.82 & 21.20 \\
		\hline
		O$_2$ & 3.71 & 21.94 & 3.59 & 26.01 \\
		\hline
	\end{tabular}
\end{table}

\


\subsubsection*{Work-function type gas sensing with applied strain}
The absolute workfunction values and their percentage changes, as compared to the strained monolayer before gas adsorption, for both tensile and compressive strains, are presented in Table~\ref{tab:wf_strain} and graphically in Fig.~\ref{fig_strain_wf}. Among the molecules NO$_2$, SO$_2$, CS$_2$, CCl$_2$O, and CO that exhibited adsorption energies exceeding the threshold of $-0.4$eV under 3\% tensile strain, NO\textsubscript{2} adsorption is characterized by a work-function change of 19.36\%. SO\textsubscript{2}  molecules result in a work-function of $4.65\%$, which is below the threshold value of $15\%$ as considered in our paper. A $14.83\%$ change in work-function for CS\textsubscript{2} adsorption is very close to the threshold work-function change of $15\%$. Thus, it can be considered as a detectable gas by work-function type sensing arrangement. The remaining molecules, that is,  CCl$_2$O, and CO result in a work-function change of $22.94\%$ and $18.21\%$ respectively when adsorbed on the surface of the monolayer. Thus, the four molecules, namely NO$_2$, CS$_2$, CCl$_2$O, and CO, in addition to NH\textsubscript{3}, NO, H\textsubscript{2}S and HCN molecules,  can be detected by work-function type sensing arrangement with the strained monolayer. Similar to the unstrained monolayer, H$_2$O, with a workfunction change of 16.49\% and adsorption energy of $-0.73$~eV with 3\% tensile strain, also shows potential for detection. Hence, the strained monolayer sensor should not be employed in humid environment unless the objective is towards moisture detection.\\ 
\indent As discussed earlier, the introduction of 2\% compressive strain does not result in any additional molecules surpassing the adsorption energy threshold of $-0.4$eV. However, for  NH\textsubscript{3},  which demonstrated adsorption energy and work-function change of  $-1.07$eV and $25.38\%$ respectively in the unstrained monolayer, now demonstrates adsorption energy of $-1.24$eV and work-function change of   $45.90\%$. Thus, under 2\% compressive strain, the detectability of the NH\textsubscript{3} molecule may improve, albeit at the expense of compromising the sensor's reusability.  Interestingly, CH\textsubscript{2}O exhibits an adsorption energy of $-0.81$,eV along with a work function change of $14.24\%$ (very close to the $15\%$ threshold work-function change) when adsorbed on the monolayer under 2\% compressive strain. It is noteworthy that CH\textsubscript{2}O, which remained undetectable by both resistive and work function-based sensing mechanisms in the unstrained and tensile-strained monolayer, becomes detectable under compressive strain via a work function-based sensing arrangement, provided a sufficiently sensitive measurement setup is employed.
\section{Conclusion} \label{conclusion}
In summary, this study comprehensively evaluated the In\textsubscript{2}O\textsubscript{3} monolayer’s sensing capabilities by investigating its interactions with ten hazardous gases—NH\textsubscript{3}, NO, NO\textsubscript{2}, SO\textsubscript{2}, CS\textsubscript{2}, H\textsubscript{2}S, HCN, CCl\textsubscript{2}O, CH\textsubscript{2}O, and CO—targeting both resistive-type and work function-based detection pathways. To ensure practical deployability, the monolayer's response to common atmospheric species, including O\textsubscript{2}, CO\textsubscript{2}, and H\textsubscript{2}O, was also thoroughly assessed, providing critical insights into its real-world deployment.  The analysis reveals that NO and H\textsubscript{2}S adsorption exhibit strong adsorption energies, substantial modifications in the electronic density of states, and significant conductivity changes, positioning them as the most promising candidates for resistive-type sensing at room temperature. Workfunction-based sensing extends the detection capability to additional analytes such as NH\textsubscript{3} and HCN, which remain undetectable through resistive sensing alone. \\
\indent Moving further, mechanical strain engineering was demonstrated as an effective strategy to enhance sensing performance and selectivity. Under moderate tensile and compressive strains, several previously weakly adsorbing gases, like NO$_2$, CS$_2$, CCl$_2$O, CH\textsubscript{2}O and CO, exhibit improved adsorption energies and increased sensing responses either by resistance based or by work-function based sensing mechanism. Although the monolayer remainds largely unresponsive to ambient gases like O\textsubscript{2} and CO\textsubscript{2},  competitive H\textsubscript{2}O adsorption underscores the necessity for controlled humidity during practical sensor deployment. Overall, the 2D In\textsubscript{2}O\textsubscript{3} monolayer offers a highly versatile platform for selective, sensitive, and potentially reusable gas sensing devices, suitable for environmental monitoring and safety-critical applications.

\section*{Author Contributions} Afreen Anamul Haque and Suraj G. Dhongade (equal contribution): conceptualization, coding, simulation, formal
analysis, investigation, data analysis and writing the original draft; Aniket Singha: conceptualization, review and data
analysis,   project administration, and funding acquisition.

\section*{Conflicts of Interest}

There are no conflicts to declare

\section*{Data Availability}
The data supporting this article have been included as part of the main text and Supplementary Information.

\section*{Acknowledgment}
A.S. acknowledges computational facility procured through financial support from the Science and Engineering Research Board (SERB) under the Department of Science and Technology (DST) via grant number SRG/2020/000593 and the Ministry of Education (MoE) through grant no. STARS/APR2019/PS/566/FS under ``STARS'' scheme. A.A.H. acknowledges the Ministry of Education, Govt. of India, for the Prime Minister's Research Fellowship (PMRF).

\bibliography{ref} 

\providecommand{\latin}[1]{#1}
\makeatletter
\providecommand{\doi}
  {\begingroup\let\do\@makeother\dospecials
  \catcode`\{=1 \catcode`\}=2 \doi@aux}
\providecommand{\doi@aux}[1]{\endgroup\texttt{#1}}
\makeatother
\providecommand*\mcitethebibliography{\thebibliography}
\csname @ifundefined\endcsname{endmcitethebibliography}
  {\let\endmcitethebibliography\endthebibliography}{}
\begin{mcitethebibliography}{20}
\providecommand*\natexlab[1]{#1}
\providecommand*\mciteSetBstSublistMode[1]{}
\providecommand*\mciteSetBstMaxWidthForm[2]{}
\providecommand*\mciteBstWouldAddEndPuncttrue
  {\def\EndOfBibitem{\unskip.}}
\providecommand*\mciteBstWouldAddEndPunctfalse
  {\let\EndOfBibitem\relax}
\providecommand*\mciteSetBstMidEndSepPunct[3]{}
\providecommand*\mciteSetBstSublistLabelBeginEnd[3]{}
\providecommand*\EndOfBibitem{}
\mciteSetBstSublistMode{f}
\mciteSetBstMaxWidthForm{subitem}{(\alph{mcitesubitemcount})}
\mciteSetBstSublistLabelBeginEnd
  {\mcitemaxwidthsubitemform\space}
  {\relax}
  {\relax}

\bibitem[Zhang \latin{et~al.}(2019)Zhang, Khan, Zou, Zhang, and Li]{b1}
Zhang,~L.; Khan,~K.; Zou,~J.; Zhang,~H.; Li,~Y. Recent advances in emerging 2D
  material-based gas sensors: potential in disease diagnosis. \emph{Advanced
  Materials Interfaces} \textbf{2019}, \emph{6}, 1901329\relax
\mciteBstWouldAddEndPuncttrue
\mciteSetBstMidEndSepPunct{\mcitedefaultmidpunct}
{\mcitedefaultendpunct}{\mcitedefaultseppunct}\relax
\EndOfBibitem
\bibitem[Buckley \latin{et~al.}(2020)Buckley, Black, Castanon, Melios, Hardman,
  and Kazakova]{b2}
Buckley,~D.~J.; Black,~N.~C.; Castanon,~E.~G.; Melios,~C.; Hardman,~M.;
  Kazakova,~O. Frontiers of graphene and 2D material-based gas sensors for
  environmental monitoring. \emph{2D Materials} \textbf{2020}, \emph{7},
  032002\relax
\mciteBstWouldAddEndPuncttrue
\mciteSetBstMidEndSepPunct{\mcitedefaultmidpunct}
{\mcitedefaultendpunct}{\mcitedefaultseppunct}\relax
\EndOfBibitem
\bibitem[Wu \latin{et~al.}(2013)Wu, Wang, Wei, Yang, and Dresselhaus]{b3}
Wu,~J.; Wang,~B.; Wei,~Y.; Yang,~R.; Dresselhaus,~M. Mechanics and mechanically
  tunable band gap in single-layer hexagonal boron-nitride. \emph{Materials
  Research Letters} \textbf{2013}, \emph{1}, 200--206\relax
\mciteBstWouldAddEndPuncttrue
\mciteSetBstMidEndSepPunct{\mcitedefaultmidpunct}
{\mcitedefaultendpunct}{\mcitedefaultseppunct}\relax
\EndOfBibitem
\bibitem[Chaves \latin{et~al.}(2020)Chaves, Azadani, Alsalman, Da~Costa,
  Frisenda, Chaves, Song, Kim, He, Zhou, \latin{et~al.} others]{b4}
Chaves,~A.; Azadani,~J.~G.; Alsalman,~H.; Da~Costa,~D.; Frisenda,~R.;
  Chaves,~A.; Song,~S.~H.; Kim,~Y.~D.; He,~D.; Zhou,~J.; others Bandgap
  engineering of two-dimensional semiconductor materials. \emph{npj 2D
  Materials and Applications} \textbf{2020}, \emph{4}, 29\relax
\mciteBstWouldAddEndPuncttrue
\mciteSetBstMidEndSepPunct{\mcitedefaultmidpunct}
{\mcitedefaultendpunct}{\mcitedefaultseppunct}\relax
\EndOfBibitem
\bibitem[Vincent \latin{et~al.}(2021)Vincent, Liang, Singh, Castanon, Zhang,
  McCreary, Jariwala, Kazakova, and Al~Balushi]{b5}
Vincent,~T.; Liang,~J.; Singh,~S.; Castanon,~E.~G.; Zhang,~X.; McCreary,~A.;
  Jariwala,~D.; Kazakova,~O.; Al~Balushi,~Z.~Y. Opportunities in electrically
  tunable 2D materials beyond graphene: Recent progress and future outlook.
  \emph{Applied Physics Reviews} \textbf{2021}, \emph{8}\relax
\mciteBstWouldAddEndPuncttrue
\mciteSetBstMidEndSepPunct{\mcitedefaultmidpunct}
{\mcitedefaultendpunct}{\mcitedefaultseppunct}\relax
\EndOfBibitem
\bibitem[Wang \latin{et~al.}(2022)Wang, Gu, Chen, Ji, Zhu, and Sun]{b6}
Wang,~B.; Gu,~Y.; Chen,~L.; Ji,~L.; Zhu,~H.; Sun,~Q.-Q. Gas sensing devices
  based on two-dimensional materials: a review. \emph{Nanotechnology}
  \textbf{2022}, \relax
\mciteBstWouldAddEndPunctfalse
\mciteSetBstMidEndSepPunct{\mcitedefaultmidpunct}
{}{\mcitedefaultseppunct}\relax
\EndOfBibitem
\bibitem[Geim and Novoselov(2010)Geim, and Novoselov]{b7}
Geim,~A.~K.; Novoselov,~K.~S. \emph{Nanoscience and technology: a collection of
  reviews from nature journals}; World Scientific, 2010; pp 11--19\relax
\mciteBstWouldAddEndPuncttrue
\mciteSetBstMidEndSepPunct{\mcitedefaultmidpunct}
{\mcitedefaultendpunct}{\mcitedefaultseppunct}\relax
\EndOfBibitem
\bibitem[Yang \latin{et~al.}(2017)Yang, Jiang, and Wei]{b8}
Yang,~S.; Jiang,~C.; Wei,~S.-h. Gas sensing in 2D materials. \emph{Applied
  Physics Reviews} \textbf{2017}, \emph{4}, 021304\relax
\mciteBstWouldAddEndPuncttrue
\mciteSetBstMidEndSepPunct{\mcitedefaultmidpunct}
{\mcitedefaultendpunct}{\mcitedefaultseppunct}\relax
\EndOfBibitem
\bibitem[Patial and Deshwal(2022)Patial, and Deshwal]{b9}
Patial,~P.; Deshwal,~M. Selectivity and sensitivity property of metal oxide
  semiconductor based gas sensor with dopants variation: A review.
  \emph{Transactions on Electrical and Electronic Materials} \textbf{2022},
  \emph{23}, 6--18\relax
\mciteBstWouldAddEndPuncttrue
\mciteSetBstMidEndSepPunct{\mcitedefaultmidpunct}
{\mcitedefaultendpunct}{\mcitedefaultseppunct}\relax
\EndOfBibitem
\bibitem[Walker \latin{et~al.}(2022)Walker, Karnati, Akbar, and Morris]{b11}
Walker,~J.; Karnati,~P.; Akbar,~S.~A.; Morris,~P.~A. Selectivity mechanisms in
  resistive-type metal oxide heterostructural gas sensors. \emph{Sensors and
  Actuators B: Chemical} \textbf{2022}, \emph{355}, 131242\relax
\mciteBstWouldAddEndPuncttrue
\mciteSetBstMidEndSepPunct{\mcitedefaultmidpunct}
{\mcitedefaultendpunct}{\mcitedefaultseppunct}\relax
\EndOfBibitem
\bibitem[Meng \latin{et~al.}(2020)Meng, Houssa, Iordanidou, Pourtois,
  Afanasiev, and Stesmans]{in2o3}
Meng,~R.; Houssa,~M.; Iordanidou,~K.; Pourtois,~G.; Afanasiev,~V.; Stesmans,~A.
  Two-dimensional gallium and indium oxides from global structure searching:
  Ferromagnetism and half metallicity via hole doping. \emph{Journal of Applied
  Physics} \textbf{2020}, \emph{128}, 034304\relax
\mciteBstWouldAddEndPuncttrue
\mciteSetBstMidEndSepPunct{\mcitedefaultmidpunct}
{\mcitedefaultendpunct}{\mcitedefaultseppunct}\relax
\EndOfBibitem
\bibitem[Grimme \latin{et~al.}(2010)Grimme, Antony, Ehrlich, and Krieg]{van1}
Grimme,~S.; Antony,~J.; Ehrlich,~S.; Krieg,~H. {A consistent and accurate ab
  initio parametrization of density functional dispersion correction (DFT-D)
  for the 94 elements H-Pu}. \emph{The Journal of Chemical Physics}
  \textbf{2010}, \emph{132}, 154104\relax
\mciteBstWouldAddEndPuncttrue
\mciteSetBstMidEndSepPunct{\mcitedefaultmidpunct}
{\mcitedefaultendpunct}{\mcitedefaultseppunct}\relax
\EndOfBibitem
\bibitem[Grimme(2006)]{van2}
Grimme,~S. Semiempirical GGA-type density functional constructed with a
  long-range dispersion correction. \emph{Journal of Computational Chemistry}
  \textbf{2006}, \emph{27}, 1787--1799\relax
\mciteBstWouldAddEndPuncttrue
\mciteSetBstMidEndSepPunct{\mcitedefaultmidpunct}
{\mcitedefaultendpunct}{\mcitedefaultseppunct}\relax
\EndOfBibitem
\bibitem[Wang \latin{et~al.}(2014)Wang, Wang, Huang, Gao, Kong, and Zhang]{b22}
Wang,~Y.; Wang,~B.; Huang,~R.; Gao,~B.; Kong,~F.; Zhang,~Q. First-principles
  study of transition-metal atoms adsorption on MoS2 monolayer. \emph{Physica
  E: Low-dimensional Systems and Nanostructures} \textbf{2014}, \emph{63},
  276--282\relax
\mciteBstWouldAddEndPuncttrue
\mciteSetBstMidEndSepPunct{\mcitedefaultmidpunct}
{\mcitedefaultendpunct}{\mcitedefaultseppunct}\relax
\EndOfBibitem
\bibitem[Haque \latin{et~al.}(2025)Haque, Dhongade, and Singha]{afreen}
Haque,~A.~A.; Dhongade,~S.~G.; Singha,~A. Predictive Analysis of Gas Sensing
  Properties in a Novel 2D Gallium Oxide Phase. \emph{IEEE Sensors Journal}
  \textbf{2025}, \emph{25}, 12644--12652\relax
\mciteBstWouldAddEndPuncttrue
\mciteSetBstMidEndSepPunct{\mcitedefaultmidpunct}
{\mcitedefaultendpunct}{\mcitedefaultseppunct}\relax
\EndOfBibitem
\bibitem[Zhu \latin{et~al.}(2022)Zhu, Xu, Ha, Li, Zhang, Zhang, and Feng]{b23}
Zhu,~J.; Xu,~Z.; Ha,~S.; Li,~D.; Zhang,~K.; Zhang,~H.; Feng,~J. Gallium Oxide
  for Gas Sensor Applications: A Comprehensive Review. \emph{Materials}
  \textbf{2022}, \emph{15}\relax
\mciteBstWouldAddEndPuncttrue
\mciteSetBstMidEndSepPunct{\mcitedefaultmidpunct}
{\mcitedefaultendpunct}{\mcitedefaultseppunct}\relax
\EndOfBibitem
\bibitem[Han \latin{et~al.}(2023)Han, Zhang, Liu, Ma, Guo, Jiang, Wan, Wang,
  Yuan, Zhou, \latin{et~al.} others]{b29}
Han,~R.; Zhang,~Z.; Liu,~W.; Ma,~F.; Guo,~H.; Jiang,~Z.; Wan,~X.; Wang,~A.;
  Yuan,~C.; Zhou,~W.; others Theoretical insights into the two-dimensional
  gallium oxide monolayer for adsorption and gas sensing of C 4 F 7 N
  decomposition products. \emph{Journal of Materials Chemistry C}
  \textbf{2023}, \emph{11}, 11928--11935\relax
\mciteBstWouldAddEndPuncttrue
\mciteSetBstMidEndSepPunct{\mcitedefaultmidpunct}
{\mcitedefaultendpunct}{\mcitedefaultseppunct}\relax
\EndOfBibitem
\bibitem[Agboola and Benson(2021)Agboola, and Benson]{b32}
Agboola,~O.~D.; Benson,~N.~U. Physisorption and chemisorption mechanisms
  influencing micro (nano) plastics-organic chemical contaminants interactions:
  a review. \emph{Frontiers in Environmental Science} \textbf{2021}, \emph{9},
  678574\relax
\mciteBstWouldAddEndPuncttrue
\mciteSetBstMidEndSepPunct{\mcitedefaultmidpunct}
{\mcitedefaultendpunct}{\mcitedefaultseppunct}\relax
\EndOfBibitem
\bibitem[Kalwar \latin{et~al.}(2022)Kalwar, Fangzong, Soomro, Naich, Saeed, and
  Ahmed]{b34}
Kalwar,~B.~A.; Fangzong,~W.; Soomro,~A.~M.; Naich,~M.~R.; Saeed,~M.~H.;
  Ahmed,~I. Highly sensitive work function type room temperature gas sensor
  based on Ti doped hBN monolayer for sensing CO 2, CO, H 2 S, HF and NO. A DFT
  study. \emph{RSC advances} \textbf{2022}, \emph{12}, 34185--34199\relax
\mciteBstWouldAddEndPuncttrue
\mciteSetBstMidEndSepPunct{\mcitedefaultmidpunct}
{\mcitedefaultendpunct}{\mcitedefaultseppunct}\relax
\EndOfBibitem
\end{mcitethebibliography}

\end{document}